\documentclass[conference]{IEEEtran}
\IEEEoverridecommandlockouts

\usepackage{cite}
\usepackage{amsmath,amssymb,amsfonts}
\usepackage{graphicx}
\usepackage{textcomp}
\def\BibTeX{{\rm B\kern-.05em{\sc i\kern-.025em b}\kern-.08em
    T\kern-.1667em\lower.7ex\hbox{E}\kern-.125emX}}

\usepackage[table,dvipsnames]{xcolor}
\usepackage[T1]{fontenc}
\usepackage{url}
\usepackage{listings}
\usepackage{float}
\newfloat{listing}{tbp}{lol}     
\floatname{listing}{Listing}
\lstdefinestyle{codestyle}{
  basicstyle=\footnotesize\ttfamily,
  numbers=left,
  numberstyle=\tiny\color{gray},
  numbersep=5pt,
  breaklines=true,
  breakatwhitespace=false,
  showstringspaces=false,
  keywordstyle=\color{NavyBlue}\bfseries,
  commentstyle=\color{ForestGreen}\itshape,
  stringstyle=\color{BrickRed},
  frame=none,
  xleftmargin=3mm,
  framesep=2mm,
  upquote=true,
}
\usepackage{mdframed}
\usepackage{capt-of}         
\usepackage{booktabs}
\usepackage{threeparttable}
\usepackage{tabularx}
\usepackage[acronym,nomain,nonumberlist]{glossaries}

\usepackage{sc26repro}

\newacronym{ai}{AI}{artificial intelligence}
\newacronym{aiml}{AI/ML}{artificial intelligence and machine learning}
\newacronym{ci}{CI}{continuous integration}
\newacronym{bsp}{BSP}{bulk synchronous parallel}
\newacronym{mpi}{MPI}{Message Passing Interface}
\newacronym{hpc}{HPC}{high performance computing}
\newacronym{aws}{AWS}{Amazon Web Services}
\newacronym{gke}{GKE}{Google Kubernetes Engine}
\newacronym{aks}{AKS}{Azure Kubernetes Service}
\newacronym{eks}{EKS}{Elastic Kubernetes Service}
\newacronym{ml}{ML}{machine learning}
\newacronym{rdma}{RDMA}{Remote Direct Memory Access}
\newacronym{os}{OS}{operating systems}
\newacronym{vm}{VM}{virtual machine}
\newacronym{llm}{LLMs}{Large Language Models}
\newacronym{fom}{FOM}{figure of merit}
\newacronym{efa}{EFA}{Elastic Fabric Adapter}
\newacronym{ec2}{EC2}{Elastic Compute Cloud}
\newacronym{ucx}{UCX}{Unified Communication X}
\newacronym{cni}{CNI}{container networking interface}
\newacronym{ebpf}{eBPF}{extended Berkeley Packet Filter}
\newacronym{gvnic}{gVNIC}{Google Virtual NIC}
\newacronym{uri}{URI}{unique resource identifier}
\newacronym{api}{API}{application programming interface}
\newacronym{dag}{DAG}{directed acylcic graph}
\newacronym{crd}{CRD}{custom resource definition}
\newacronym{mcp}{MCP}{Model Context Protocol}
\newacronym{pid}{PID}{Process IDentifier}

\newcommand{\yes}{\cellcolor{LimeGreen!70} Yes}

\begin{document}

\title{Hierarchical Server Architecture for Agentic Science}


 \author{\IEEEauthorblockN{Vanessa Sochat}
 \IEEEauthorblockA{\textit{Lawrence Livermore National Laboratory} \\
 Livermore, California, USA \\
 sochat1@llnl.gov \\
 ORCID 0000-0002-4387-3819}
 \and
 \IEEEauthorblockN{Daniel Milroy}
 \IEEEauthorblockA{\textit{Lawrence Livermore National Laboratory} \\
 Livermore, California, USA \\
 milroy1@llnl.gov \\
 ORCID 0000-0001-6500-3227}
 }

\maketitle

\begin{abstract}
Agentic science is transforming the landscape of computational work, and is applied to scientific pipelines and workload managers. Scientific workloads require specialized hardware within and between institutions. Automated resource discovery is an essential step for scheduling workloads with specific hardware and environmental requirements. In this paper, we present a hierarchical, dynamic architecture and accompanying software to discover resources across diverse cloud, edge, and HPC systems. The design enables concurrent, asynchronous negotiation, selection, and dispatch of requests for work using secretary agents. The agents probe and discover 51 real and simulated providers across 7 categories. We perform 19,973 negotiation and 6,952 selection simulations to assess reliability of decisions, demonstrating high (87.71\%) negotiation accuracy and selection costs comparable to more traditional strategies. Designed for extensibility and currently supporting the US DOE Genesis Mission, this architecture exemplifies the importance of careful coordination between agents, discovery tools, and infrastructure for agentic science.
\end{abstract}

\begin{IEEEkeywords}
agentic science, resource discovery, Model Context Protocol, multi-agent systems,
converged computing, workload dispatch, negotiation, HPC
\end{IEEEkeywords}

\section{Introduction}

The high performance computing (HPC) center of the future will not just be a facility for running jobs. It will be a proactive, autonomous platform that accelerates science. A central component of this setup will be \gls{aiml} models that can receive and intelligently respond to requests. In this possible future, a user no longer submits a job as a structured script with directives for a specific workload manager. It is provided as a textual request for resource quantity and type, and a preference for priority, time to completion, cost, and performance. Work in converged computing \cite{sochat2024converged} provides a foundation for how traditional HPC technologies can be integrated with and extended to cloud-like environments \cite{sochat2025usability}. The movement serves as a means to transition from decades of HPC-specific techniques to more flexible practices that integrate with a broader range of environments. 

The converged computing movement encompasses work related to scheduling \cite{misale2021towards,wu2024research,hazlewood2025integration,milroy2022}, workload orchestration \cite{sochat2025state}, storage technologies \cite{badia2017workflows}, and network \cite{jolicoeur2025enabling,link2019container}. Converged computing can extend from specific environments to an entire HPC center, including not just automation, modularity, and declarative center management, but also integration of \gls{llm} and agents to accomplish scientific tasks. While early work that brought \gls{llm} into science focused on hypothesis generation and literature review, it has now expanded to encompass experimental design, application optimization \cite{10.3389/frai.2025.1649155}, and domain-specific analysis \cite{Xin2025-il}. However, an equally important task is extending agentic intelligence to resource provisioning and workload management. National incentives such as the Genesis Mission \cite{Gil2025Genesis} are laying the foundation for infrastructure and software based around agents.

There is ample existing work on enabling distributed workflows in cloud environments in multi-cluster environments~\cite{wang2025efficient,bahreini2024caspian,el2026survey}. A common strategy is to dispatch a submitted job to all or multiple clusters in a set, and the first cluster to run is given ownership \cite{federated-slurm,kueue}. This approach assumes that running tasks as soon as possible is the preferred criterion for cluster selection. However, a resource available sooner may not be ``best'' for the work. A resource available later may allow an application to execute faster, or have a better suited resource specification or lower cost. 

There is a distinction between resource provisioning and workflow management. While both account for application-specific needs, the first falls under the domain of administration, hardware, and dispatch, while the second is focused on application parameter selection, step coordination, and decision making. An \gls{hpc} center is no longer a well-understood environment that can be expected to have a specific workload manager or software suite. It could consist of a dynamic Kubernetes cluster running on cloud resources, a set of virtual machines with Slurm, a single server providing access to an edge device, or a traditional on-premises resource. In addition, a strategy for dispatching work cannot assume a specific manager, software distribution strategy, or network. The resources must be probed, discovered, and made accessible not just to a local resource manager, but potentially to a fleet --- a managed group of compute resources that serves a broad set of users, use cases, and workload types. A fleet generally has a unified control plane, federated job scheduling, heterogeneous resource management, and multi-cluster data orchestration.

In this work, we present a novel architecture and supporting software to enable probing, discovery, proposal generation, and selection of workloads. We make the following contributions:

\begin{itemize}
\item{Implementation of agentic hierarchical server architecture}
\item{19,973 experiments to assess reliability of job negotiation}
\item{6,952 simulations for testing of job selection}
\item{51 simulated and real resource provider interfaces}
\item{Resource secretary software/agents for provider discovery}
\item{Algorithms for prompt and worker archetype generation}
\item{Over 20K open source datasets with detailed interactions}
\end{itemize}

In Section \ref{section:mcp-server} we describe the server and software components. The server is multi-faceted, serving as a unit in any level of a hierarchy of servers -- a hub, worker, both, or neither -- to be deployed under a job, at the top of a cluster, or a fleet. The server as a standalone unit can provision tools, event streams, and standard prompts and resources for agents. In a hierarchy, communication can propagate asynchronously and concurrently between child and parent levels, allowing for intelligent and agentic job submission and dispatch. Agentic secretary and discovery software expose real and simulated resource providers, allowing a worker to provide on-demand information or status. The architecture supports credential-based registration to join a parent and respond to work requests. In Section \ref{section:resource-secretary} we describe the agentic secretary that handles job negotiation, selection, and dispatch. We perform 19,973 and 6,952 negotiation and selection simulations (Section \ref{sec:experiments}) with 200 worker archetypes for \gls{hpc}, cloud, and edge devices to assess prompts generated from a gradient of specificity. Our negotiation agents are able to discover simulated providers and give an accurate satisfiability assessment -- a decision of whether a computational resource can meet the needs of a job (accuracy= 87.71\%). Our selection agents show cost efficiency comparable to more traditional strategies. Furthermore, we demonstrate the architectural capabilities to orchestrate job negotiation experiments at several levels. We present practical advice for design of descriptive, agentic dispatch systems and conclude with discussion (Section \ref{sec:discussion}) about lessons learned and future work.

\section{Methods}
\label{sec:methods}

\subsection{Model Context Protocol Server}
\label{section:mcp-server}

The \gls{mcp} was developed by Anthropic to standardize LLM interactions with a server. It was donated to the Agentic AI Foundation in 2025, and is now an industry standard \cite{mcp} that provides a defined set of json-rpc calls expected between a client and server. 

\begin{figure}[ht!]
    \centering
    \includegraphics[width=0.53\textwidth]{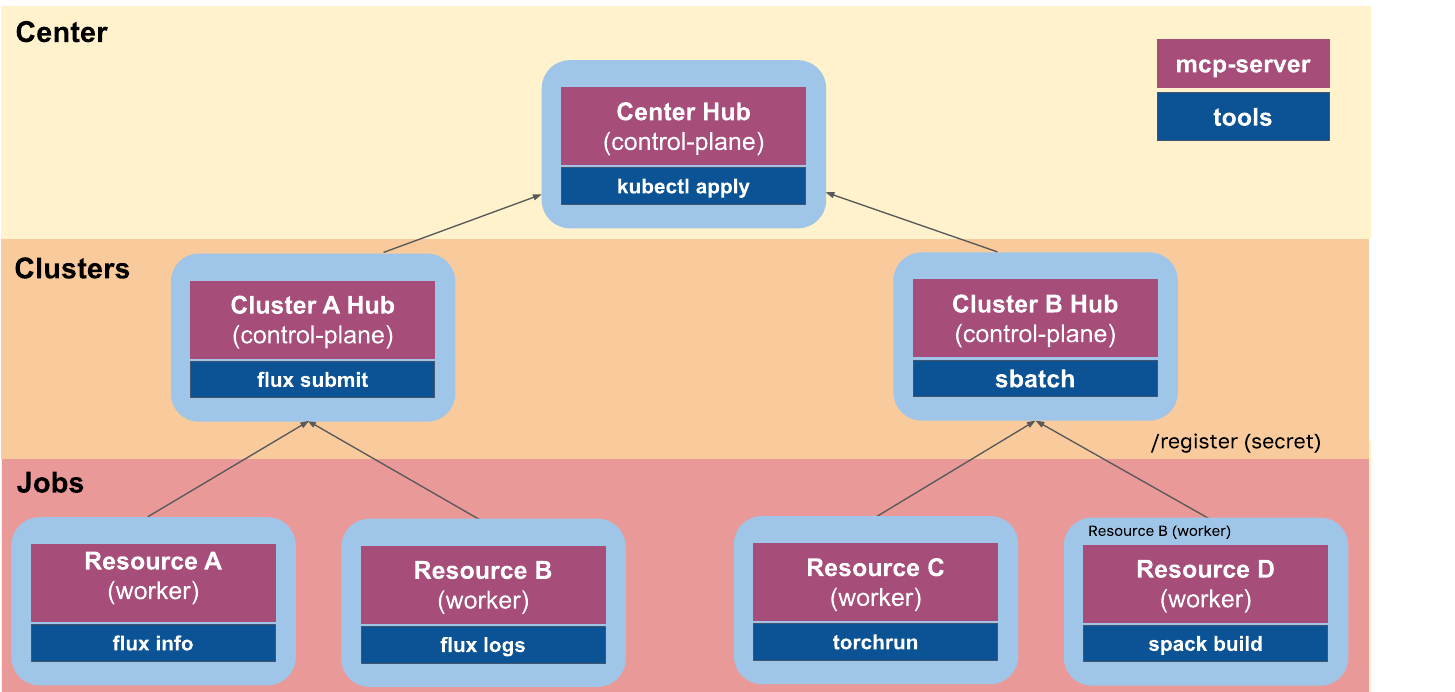}
  \caption{\textbf{The mcp-server hierarchy at multiple levels}. When deployed under the context of a \emph{Job}, the server exists for the lifecycle of the job and is user owned and provisions functions for an application or workflow orchestration. At the level of the \emph{Cluster} it requires multi-tenancy authentication and authorization and runs as a persistent service. On the level of a \emph{Center}, the server requires the highest level of security. Each child server must register to the fleet with a secret.}
  \label{fig:mcp-hierarchy}
\end{figure}

The model context protocol server (mcp-server) project \cite{vanessa_sochat_2026_19476302, mcpserver} aims to provide a server for the dynamic addition of resources to jobs, cluster services, and orchestrations that need to serve LLM agents. A user can dynamically add tools, prompts, resources, and events at the level of a single node, a job, a shared cluster resource, or a center by way of a YAML configuration file (Figure \ref{fig:mcp-hierarchy}). A parent worker that distributes requests to child workers is considered a hub. A group of computational workers is the fleet. The requests can be batched, serial, or done in parallel. Any computational resource can start a server with an address and credentials for a parent to register to a larger network. For the purposes of this work we consider workers as cloud, edge, and on-premises clusters, either real or simulated. 

\smallskip
\noindent{\bf Hierarchical Design} 
\label{section:hierarchical-design}
An mcp-server can be deployed at several levels that map to use cases for an application, cluster, or center (Figure \ref{fig:mcp-hierarchy}). The hub is designed to minimize computational work, and distribute requests efficiently using asynchronous, concurrent calls. Its primary role is to receive and transmit requests. Computational work such as serving models for \gls{llm} or executing decision algorithms are placed on leaf workers or the calling client. While a hub can be included in a fleet and dually act as a worker, we recommend a standalone hub for production use cases.

\smallskip
\noindent{\bf Authentication and Authorization} 
\label{section:authz-authn}
An mcp-server hub can be deployed with or without SSL or authentication and authorization (authN and authZ). These interfaces are provided as middleware, and can be extended to include institution-specific needs. Token-based authentication is provided for a default. Current national initiatives focused primarily on authZ and authN, the detailed design which is out of scope for this work  \cite{Gil2025Genesis,llnl_fghpcc_rfi_2025}. For our local testing and simulations we did not enable authN and authZ.

\smallskip
\noindent{\bf Capabilities} 
\label{section:capabilities}
The \gls{mcp} standard defines capabilities as tools (executable functions), resources (read-only data), and prompts (templates for agents). While it is common practice to see libraries for agentic science implement servers and tools alongside one another with hard-coded decorators for discovery \cite{menon2026multi, ursa}, the mcp-server takes a more flexible approach by loading capabilities dynamically from a configuration file along with custom capabilities that a specific application might need at execution time. 

The mcp-server adds event subscriptions to extend traditional \gls{mcp}. We provision a Subscription Manager that can register, validate, and deliver events from a local provider using HTTP notifications \cite{notification-json-rpc-2}. A user server can register an event streaming class (e.g., events from workload managers Kubernetes and Flux \cite{ahn-2014}) on the fly, and events are collected in the background that can be delivered as they come, or in batches between looped requests.

\begin{figure}[ht!]
    \centering
    \includegraphics[width=0.8\textwidth]{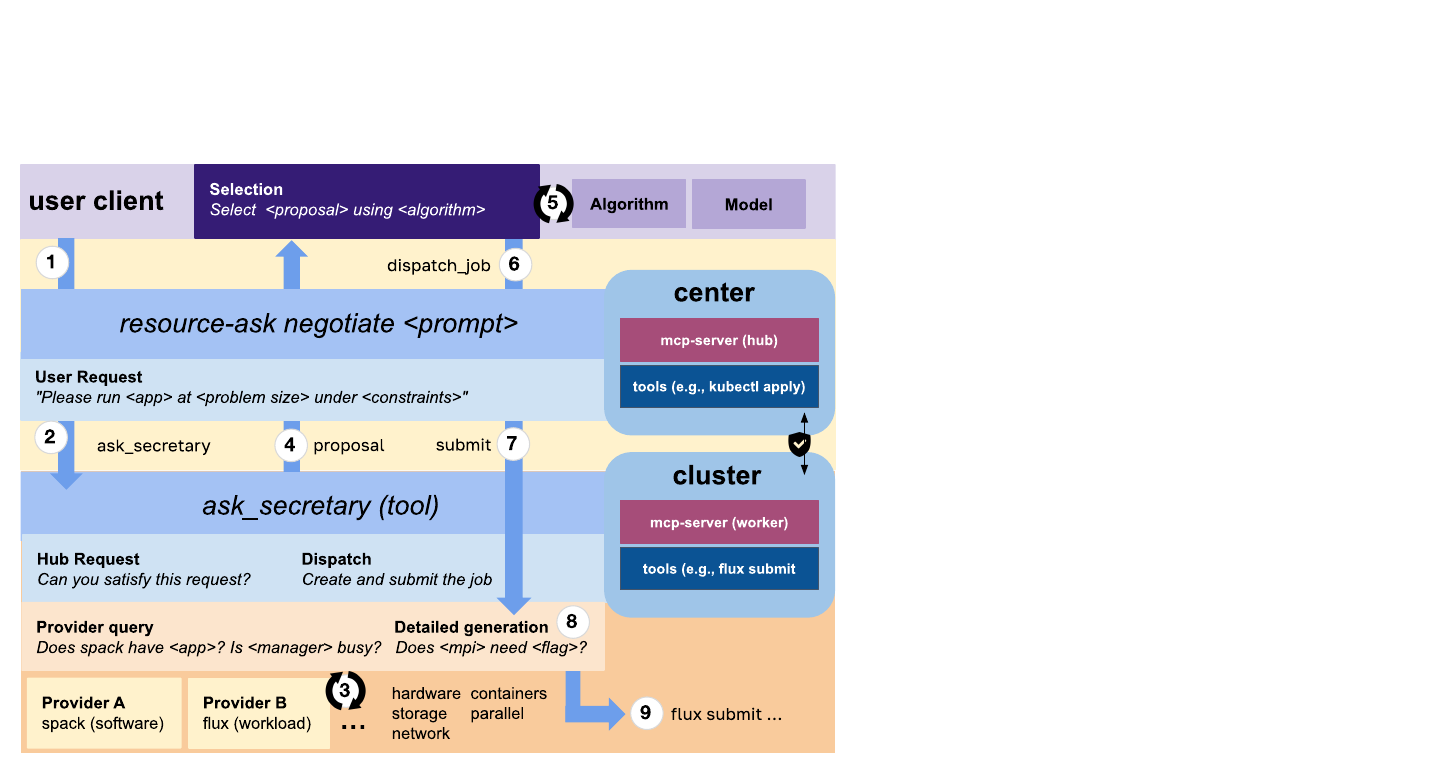}
  \caption{\normalfont \textbf{Job Negotiation}. The client makes a request for an application, resources, and constraints (1) received by the hub that makes a concurrent request to workers via \emph{ask\_secretary} (2). The secretary agent of each cluster receives the request. Resource providers allow for dynamic query (3). The agent explores until it has sufficient information to return a proposal (4). A selection algorithm is used (5) to choose a cluster. The client calls \emph{dispatch\_job} to dispatch the work to hub (6), and the hub submits (7) to the selected cluster. The secretary agent transforms the request into a job for the workload manager (8) and the work is executed (9). }
  \label{fig:job-negotiation}
\end{figure}

\smallskip
\noindent{\bf Job Negotiation} 
\label{section:job-negotiation}
When running as a hierarchy, the hub and workers handle job negotiation. Job negotiation is the receipt of a textual request for work that leads to dispatch to a specific computational resource in the fleet (Figure \ref{fig:job-negotiation}). The steps for job negotiation include:

\begin{itemize}
\item \emph{Submit} a textual request for work (application and resources).
\item \emph{Query} the fleet to evaluate if the request can be satisfied.
\item \emph{Receive} proposals from distributed secretary agents.
\item \emph{Select} a proposal based on algorithm and user policy.
\item \emph{Dispatch} work to the target cluster.
\end{itemize}

We provide software, the ``Resource Secretary," that provides a client to easily perform the interactions above and a Secretary Agent (Section \ref{section:secretary-agent}) that evaluates the computational resource for satisfiability. The software enables resource provider discovery for the worker (Section \ref{section:provider-selection}). We discuss job negotiation next.

\section{Agentic Negotiation}
\label{section:agentic-negotiation}

\smallskip
\noindent{\bf Workload Submission} 
\label{section:workload-submission}
The initial steps for negotiation consist of querying a set of resources  to determine work satisfiability and receiving back proposals from workers. A user submits a textual request for work to a hub to the \emph{negotiate\_job} endpoint. The prompt can take on any level of detail that is desired, for example, including a workload and scale to run, policy or priority for the job, and needs for time and cost. The request can range from an exact command to a generic description that the agent must parse. If an agent does not have enough information to complete a request during negotiation, it returns the request with feedback to the user. The flow of a request is shown in Figure 2. The request is received by the hub, and is parsed into a prompt and sent out to workers asynchronously and concurrently using a gather function \cite{asyncio-gather}. A worker receives the request via the \emph{ask\_secretary} function, where a Secretary Agent (Section \ref{section:secretary-agent}) is instantiated to serve as a representative for the computational resource. By way of a discoverable provider interface, the agent is able to see and query the current state of the resources it owns. Each \emph{SecretaryAgent} instance is scoped to one job request for security -- information cannot be shared across jobs. Computational resources can use private or on-premises models. The model name, type, and credentials are discovered in the environment, and the model connection is created by a generic interface that handles the OpenAI \gls{api} standard and Google Gemini. This interface can be extended to other \gls{api}s. Secretary Agents respond to the calling server with proposals for work.

\smallskip
\noindent{\bf Selection} 
\label{sec:selection}
Secretary Agent proposals return from workers and are received by the parent hub. The parent hub immediately sends them back to the calling client to perform selection. Selection is the process of choosing a cluster from the set of proposals. The selection algorithm is run by the calling client, removing any computational burden from the fleet or a hub, and allowing use of private models or custom algorithms and interfaces written by developer users. The client uses the resource secretary software selection algorithm interface to call one or more selection algorithms (Table \ref{table:selection-algorithms}) to choose a cluster. These algorithms currently are a combination of those common to \gls{hpc} (N=4) and one agentic. For experimental use, the secretary agent command line tool can also receive proposals that have been saved as JSON to perform the same selection. 

A single proposal carries a verdict of \emph{READY} (the cluster can satisfy and receive the work immediately), \emph{BUSY} (the cluster can satisfy and receive the work later), or \emph{INCOMPATIBLE} (the cluster cannot satisfy the needs of the job). During negotiation, the agent must provide reasoning along with metrics and constraints in a JSON formatted response. These statuses are primarily provided to assist the selection algorithms described in Table \ref{table:selection-algorithms}. If no proposals can accept work, the user receives feedback, and can update and resubmit the request.

\smallskip
\noindent{\bf Workload Dispatch} 
\label{section:agentic-dispatch}
A selected proposal is sent as a dispatch request to the hub via the \emph{dispatch\_job} endpoint, and the work is sent to the Secretary Agent representing the chosen cluster, again as a prompt. The agent takes ownership of the job, using a prompt informed by its local resources along with dispatch tools provisioned by the providers to submit the work. Ownership is not just preparation of the submission, but also monitoring state update. Toward this aim, while not required for a  worker, it is suggested to use streaming events (Section \ref{section:mcp-server}) for this purpose.

\subsection{The Resource Secretary} 
\label{section:resource-secretary}
The resource secretary software \cite{vanessa_sochat_2026_19476501, resource-secretary} is a generic interface library that defines 51 real and simulated resource providers (Table \ref{table:resource-providers}) for use by an LLM-driven Secretary Agent to determine the appropriateness of a work request for a computational resource. The library has a command line interface, \emph{resource-ask}, to expose job negotiation or component steps.







\smallskip
\smallskip
\noindent{\bf Provider Interface} 
\label{section:provider-interface}
Providers use a common underlying interface to call \emph{probe} against a system, and provision a \emph{metadata} property with the provider name, description, and function argument signature for an agent to use. Decorators \emph{secretary\_tool}  and \emph{dispatch\_tool} identify class functions for negotiation and dispatch agents, respectively. Each decorated function exposes itself to the agent via parsing of inputs, outputs, and docstrings. At the time of negotiation or dispatch, a function manifest is provided to the Secretary Agent along with instructions for call formatting. As an example, a request that requires LAMMPS might proceed as follows. The Secretary Agent is instantiated, and the \emph{SpackProvider} probe detects \emph{spack} via the \emph{SPACK\_ROOT}. The agent sees that Spack is a package manager with a tool \emph{find\_package} that is documented to search for software of interest. The agent calls the function, learns that LAMMPS is installed as a package, and continues with other tool calls until a decision can be made. 

A provider is any resource that might be discovered by a worker, for example workload managers, hardware (e.g., CPU, GPU, memory), storage, distributed interfaces like the \gls{mpi}, container technologies, software package managers, and network. The worker automatically probes for all provider types, and cluster administrators can choose to selectivity disable a provider, or better expose a provider (e.g., \emph{spack} can also be discovered if the \emph{spack} executable is on the path). Joining a hierarchy does not require a cluster to expose any or all resources. Providers are designed to be read-only, and expose state data about density, quantity, and type. The provider interface is simple for developers to easily add or request custom resource providers. Simulated provider interfaces are also available to run experiments to understand each step in the negotiation process, and agent design (Section \ref{section:simulated-provider}).

\smallskip
\noindent{\bf The Negotiation Agent} 
\label{section:secretary-agent}
The \emph{SecretaryAgent} \emph{negotiate} function prepares a prompt to instruct the agent how to respond to a negotiation request with a proposal. First, the agent must analyze the user's request against provider manifests to determine if the cluster can meet the needs of the job. A provider manifest (Section \ref{section:provider-interface}) informs the agent of provider tools available, each of which is a controlled call to an instantiated class that runs in the same process. The agent can format one or more calls to learn more about the environment within one interaction using the LLM. Negotiation requires the agent to make at least one observation of the system. The agent is instructed never to provide generic chat, explain how to use software, or misrepresent the resource. The agent must also determine if the cluster is compatible based on state, and verify that software is installed and resources are sufficient. We require the agent to make at least one observation of the system, and return a response with a verdict that can be received and parsed by the calling client, as described in Section \ref{section:job-negotiation}.

\smallskip
\noindent{\bf The Dispatch Agent} 
\label{section:dispatch-agent}
Upon receiving a dispatch request from the hub, the \emph{SecretaryAgent} of the selected computational resource receives a \emph{submit} request to finalize dispatch of the work. The agent is first tasked with transforming the textual request into a job for the workload manager on the computational resource it serves. The \emph{SecretaryAgent} transforms the textual request into a job specification, and further queries the system via providers to fine tune it. Functions decorated with \emph{dispatch\_tool} are provided as a manifest to this agent for use to submit jobs and further query. In practice, the generation of the job specification means that the agent will discover exact software or library paths, and generate a command that includes application flags, environment, and workload manager parameters. When the job specification is ready, the agent executes provider tools to submit and query status. It can return updated state to the hub via event streams.

\begin{table}[htbp] 
 \centering
 \begin{threeparttable}
 \small
    \caption{Resource Secretary Selection Algorithms}
    \label{table:selection-algorithms}
     \begin{tabularx}{\columnwidth}{@{} l l l X @{}}
        \toprule
        \textbf{Strategy} & \textbf{Logic Type} & \textbf{Status} & \textbf{Primary Use Case} \\
        \midrule
        agentic & LLM Reasoning & Varies & Handles complex requests \\
        first-ready & Greedy & READY & First available with \emph{READY} status \\
        random & Stochastic & READY & Random selection \\
        soonest & Quantitative & R/B\tnote{a} & Shortest \textit{ETS}\tnote{b} \\
        run-anytime & Stochastic & R/B & All compatible clusters \\
        min-cost & Economic & R/B & Budget optimized \\
        queue-depth & Quantitative & R/B & Fleet-wide load balancing \\
        \bottomrule
    \end{tabularx}
\begin{tablenotes}[flushleft]
    \footnotesize 
    \item \textbf{a)} READY or BUSY status, with preference to READY \hspace{0.2cm} \textbf{b)} Estimated time to start. Cost is computed per-cluster by multiplying unit costs of simulated resources by counts.
\end{tablenotes}
 \end{threeparttable}
\end{table}

\subsection{Experimental Simulations}
\label{sec:experiments}

To assess the reliability and functionality of the server architecture, we performed simulations to test negotiation and selection components. The simulations are based on the logic that we can generate simulated workers according to computational resource archetypes, control random generation of the state of the resources they provision, and then assess the reliability of the agents to report it accurately. The simulations can easily be run on a single machine, giving insights to agent behavior and ability. We will perform simulation experiments to test each of negotiation and selection. For negotiation experiments, we will deploy a standard mcp-server and generate a fleet of 200 simulated workers as background processes in a different terminal (Section \ref{section:simulated-workers}). A subset of responses and metadata for 100 workers will be carried forward from negotiation simulations for selection experiments. 

\smallskip
\noindent{\bf Simulated Workers} 
\label{section:simulated-workers}
The simulation experiments start with negotiation. The hub and fleet of 200 simulated workers is deployed using the \emph{--mock} flag to start each worker. Each worker conforms to and creates simulated providers based on a computational resource archetype (Section \ref{section:provider-selection}) at a breakdown of 40/40/20 for \emph{\gls{hpc}}, \emph{cloud}, and \emph{standalone}, respectively.  To drive generation of the provider state, scale, and density, a lower level configuration is provided upon generation that is oriented to an archetype of interest. A prompt generation strategy that handles different levels of complexity and style across 10 different resource dimensions will allow us to study the impact of increasing levels of complexity on agent accuracy. A provider can be a member of one or more archetypes. While some providers are shared between spaces, others tend to be unique to the space. If we create a randomized, simulated worker, we want it to be representative of a likely archetype.

\smallskip
\noindent{\bf Worker Generation via Archetype} 
\label{section:provider-selection}
A new worker comes up and selects an archetype based on the distribution stated in the previous paragraph. We use a procedural generation to create 100 deterministic and auditable prompts to give to each of 200 of our simulated workers, for a total of 20,000 potential tests. We chose to run the simulations in serial to guarantee no violation of request limits, and to allow for execution on a single machine and careful observation. Each prompt is generated by a simulation module provided by the Resource Secretary library that can receive an archetype that drives the density and scale of generation.

Worker identity is used to start the process. The worker identifier (e.g., ``10") is hashed into an integer to better reproduce the state. Using the seed, the generator performs a weighted random selection to assign one of three archetypes: \emph{cloud}, \emph{\gls{hpc}}, and \emph{standalone}, at the breakdown of 40/40/20, respectively. Cloud generally represents the most heterogeneous of the setups, testing a secretary agent's ability to handle variety and non-standard naming conventions. The \gls{hpc} archetype brings in traditional and specialized providers that are often strictly managed, and  standalone is a long tail that encompasses the gamut from virtual machines to serverless and edge-computing.  Archetype generation logic is shown in Table \ref{tab:archetypes}.

\begin{table}[hb!]
\centering
\caption{Simulated Worker Archetypes}
\label{tab:archetypes}
\small
\begin{tabularx}{\columnwidth}{@{}X@{} l l X@{}} 
\toprule
\textbf{Arch.} & \textbf{Distr.} ($\mu, \sigma$) & \textbf{Cat.} [choose] & \textbf{Provider Slots} \\ \midrule
\textbf{HPC} & $T_s: (0.7, 0.2)$  & Workload [1] & Flux, Slurm \\
(40\%) & $T_d: (0.8, 0.2)$ & Software [1-3] & Spack, Modules, Conda, Pip \\
& & Storage [1-2] & Lustre, NFS \\
& & Network [1] & InfiniBand, OmniPath \\
& & Parallel [1-2] & OpenMPI, MPICH \\
 & & Hardware [1] & Hardware \\
   & & Container [1] & Singularity, Podman \\ \midrule
\textbf{Cloud} & $T_s: (0.3, 0.2)$ & Workload [1] & Kubernetes \\
(40\%) & $T_d: (0.4, 0.2)$ & Software [1,2] & Conda, Pip \\
 & & Storage [1-2] & S3, NFS \\
 & & Network [1] & Ethernet \\
 & & Hardware [1] & Hardware \\
 & & Container [1] & Docker \\ \midrule
\textbf{Standalone} & $T_s: (0.3, 0.2)$ & Workload [1] & Machine \\
(20\%) & $T_d: (0.4, 0.2)$ & Software [1-2] & Conda, Spack, Pip \\
 & & Storage [1] & Local-scratch \\
 & & Network [1] & Ethernet \\
& & Hardware [1] & Hardware \\
 & & Container [0-2] & Docker, Singularity \\ \bottomrule
\end{tabularx}

\smallskip
\begin{minipage}{\columnwidth}
\footnotesize \textit{$T_s, T_d$ represent latent Scale and Density targets sampled from $\mathcal{N}(\mu, \sigma)$ and clipped to $[0.01, 1.0]$. Bracketed numbers indicate slot cardinality (min-max). The hardware provider provisions each of CPU, GPU, and memory, with the potential for generating clusters without any GPU. Hardware always include CPU, GPU, and memory.}
\end{minipage}
\end{table}

Each archetype defines a scale $T_s$ and a density $T_d$ target, which provide instruction for generation of a resource based on scale (size) or density (complexity). For each, we specify the character of a worker by sampling two latent variables from a broad Gaussian distribution $\mathcal{N}(0.5, 0.2)$, mapped to $[0.01, 1.0]$, where both the mean and standard deviations can be customized by archetype or simulated provider. While each provider defines detailed metadata, the archetype gives ranges for prominent features of interest. Ranges for HPC (100-10K nodes with 32--128 CPUs per node, 128-1024 GB RAM per node, 0-8 GPUs, and 10TB-10PB storage, 3-10 partitions, 50-500 applications), Cloud (1-500 nodes with 2--64 CPUs per node, 4-256 GB RAM, 0-8 GPUs, and 10GB- 5TB, 1-2 partitions, 10-50 application), and standalone systems (1 node, 4-32 CPUs, 8-128 GB RAM, 0-1 GPU, 100GB-2TB local storage, 20-100 packages) reflect these preferences.

A large scale value (e.g., 0.9) would mean a specific provider resource is large (e.g., thousands of nodes, or many GPUs). A high density would indicate a high degree of variety, such as a large number of types of Conda packages, or workload manager partitions. It is up to each provider to receive the archetype and use the provided functions to generate counts for resources of interest. For example, an \emph{HPCArchetype} might be created that holds targets ($T_s, T_d$) that define mean ($\mu$) values for future generation. Two different instances of the same archetype can hold slightly different means, however they are sampled from the same range for the archetype. A \emph{MockSlurmProvider} might then receive the archetype, and need to generate a state relevant to scale like \emph{node\_count}. The instance of the mock provider will sample from a second Gaussian, $\mathcal{N}(T_s, 0.05)$, that accounts for the archetype scale to determine an actual node count. The value of 0.05 is a parameter that a specific provider can use to define how tight the sampling is. As an example, a worker with a scale target of 0.8 might use the target to generate 8,100 nodes and 7,900 TB of storage. An example is shown in Code \ref{code:generation}. 

For provider selection, each archetype defines a slot for a subset of relevant resource types (e.g., compute or storage), and a cardinality for each. For example, a \emph{HPCArchetype} might allow 1-3 software package managers, and select from a set of 3. The archetype is a blueprint to define rules for generation of the worker and providers. The final simulated worker is a realistic representation of a kind of computational resource that was procedurally generated using stated statistical distributions.

\smallskip
\noindent{\bf Simulated Providers} 
\label{section:simulated-provider}
A simulated provider is a mock of a real provider that is populated by an archetype, and is configured to determine scale and density of a resource. Real providers with paired classes that afford simulation are shown in Table \ref{table:resource-providers}. The selection of a subset of real providers reflects the needs of the simulation experiments. For example, a simulated prompt from a user might ask for a cluster with a low latency network, a rootless container technology, or GPU count without needing to disambiguate vendor.

\begin{listing}[ht]
\begin{mdframed}[linecolor=white, topline=true, bottomline=true, leftline=false, rightline=false, backgroundcolor=white]
\begin{lstlisting}[style=codestyle, language=Python]
class MockProvider:
    def probe(self):
        # Allow drift of v1, tight selection v2
        self.generate("v1", mode="scale", volatility=0.4)
        self.generate("v2", mode="density", volatility=0.05)
\end{lstlisting}
\captionof{listing}{\normalfont Example generation that happens during probe. Volatility is the standard deviation.}
\label{code:generation}
\end{mdframed}
\end{listing}

\smallskip
\noindent{\bf Large Language Model} 
\label{section:large-language-model}
We chose to use Google Gemini and the Python SDK, and specifically Gemini 2.0 Flash \cite{Pichai2024-sp} to optimize efficiency, speed, and cost-effectiveness (10 cents/1 million requests). The context window is 1 million tokens and allows us to run scaled experiments with reasonable cost for large numbers of requests. We use the \emph{--serial} flag provided by the hub start command to avoid exceeding rate limits and to facilitate observation of the experiments. 

\smallskip
\noindent{\bf Prompt Generation} 
\label{section:prompt-generation}
To evaluate the agentic reliability, we used dimensional sampling of three different prompt styles across 7 different specificity indices (\emph{SI}), where each index is an embedded dependent clause that incrementally adds features to increase the complexity of the request. The \emph{SI} represents a logical AND/OR gate that the agent must pass to reach a conclusion. We randomly select each SI with weights of 5, 20, 30, 20, 10,10 and 5, for 1-7, respectively, to get a distribution with fewer under-specified or highly specified prompts. Attributes include resource names, counts, and versions, along with specific environment features and temporal needs. An application must be requested, and for \emph{SI} 2 or greater, we request a version with a randomly selected operator in \texttt{<=}, \texttt{>=}, or \texttt{==}. For example, a prompt in the experiment with a specificity of 3 might be ``I want a system to run mixbench (>= 1.0.9) with 16 cores with support for low latency MPI.'' The SI level requires an application, compute specification for CPUs and optionally GPUs, and parallel software. The values for each possible attribute are selected from a global catalog of features that are extracted as the unique values and ranges across all simulated workers. Knowing the ground truth of each prompt allows us to programmatically assess agent accuracy and discovery, and determine the degree to which an answer was thoroughly investigated versus a hallucination or guess.

\begin{table*}[htbp]
 \centering
 \begin{threeparttable}
 \small
  \caption{Resource Providers Provided by the Resource Secretary Library}
  \label{table:resource-providers}
  \begin{tabular}{@{}l p{0.40\textwidth} p{0.34\textwidth}@{}}
      \toprule
        \textbf{Category} & \textbf{Simulated and real} & \textbf{Real only} \\
        \midrule
        Workload & Flux, Kubernetes, Machine, Slurm & Cobalt, Moab, Oar, Pbs, Torque \\
        Software & Conda, Pip, Modules, Spack & --- \\
        Container & Docker, Podman, Singularity & Charliecloud, Shifter \\
        Storage & Local-scratch, Lustre, Network-fs & Beegfs \\
        Network & Ethernet, Infiniband, Omni-path & --- \\
        Hardware & AMD GPU, CPU, Memory, NVIDIA GPU & --- \\
        Parallel & Mpich, Openmpi & Spectrum-mpi \\
  \bottomrule
\end{tabular}
\begin{tablenotes}[flushleft]
    \footnotesize
    \item The \emph{resource-secretary} library provides a provider interface with providers across 7 categories. Providers in the \textbf{Simulated and real} column can be driven by simulated state as well as probed on real systems; those in \textbf{Real only} are probed on real systems.
    \item Addition of a new provider uses a common base class, and only requires a probe and discovery functions.
\end{tablenotes}
 \end{threeparttable}
\end{table*}

\smallskip
\noindent{\bf Audit and Analysis} 
\label{section:audit}
The Resource Secretary comes with an Auditor that is designed to handle a worker's truth paired with a verdict and tool calls, and perform a full trace and assessment for accuracy. The process starts with comparison of the worker's ground truth to the needs of the prompt to determine if the simulated cluster could actually satisfy the work. We then assess the agent verdict for correctness. We generate a list of minimum tool calls that would be required to come to the final decision, and compare to the agent's actual calls. In the case that a request is truly \emph{INCOMPATIBLE}, for example, we cannot require a complete exploration of the entire space as the agent can return as soon as it sees the missing requirement. Instead, we allow for a smaller, still definitive subset. This process allows us to disambiguate a rigorous discovery process from hallucinated success. We assess outcomes across complexity and styles of prompts. To reflect the specificity index of the prompt, where the index reflects an incremental increase in number of features requested, we calculate a more exact score for each prompt by counting the number of required features.  For tracing tool calls, we calculate a final score as $(satisfied)/(required)$ for tool categories. For example, if an agent is required to explore the software and network subsystems to respond to a prompt but only looks at software, the score would be $1/2$ or 0.5.

\smallskip
\noindent{\bf Selection Simulation} 
\label{section:selection-simulation-method}
We are interested to assess the ability of an agentic strategy to perform selection against 6 traditional heuristic strategies (Table \ref{table:selection-algorithms}). We will run 10 full simulations for each selection algorithm (Table \ref{table:selection-algorithms}). The algorithms for \emph{first-ready}, \emph{random}, \emph{soonest}, \emph{run-anytime} are considered base cases, and \emph{min-cost} and \emph{queue-depth} are considered more informed approaches. For the latter, we will adjust the agent prompt to not only include a summary of cluster state, but also per-unit cost data. For example, to compare the agentic strategy to the \emph{min-cost} strategy, we will include per unit price data in the prompt, and a calculated total cost to the native algorithm. For each simulation, we start with a subset of 10K results across simulated workers and filter down to those that are actually compatible, as determined by the worker truth. We will run two types of simulation. The first will start workers with simulated static state and update when requests are selected. The second will assume a consistent static state for each prompt request across workers. In both cases, we are primarily interested in the agent's ability to improve upon resource efficiency and cost.

We format the result data from our negotiation experiments into a single JSON file for each prompt that was satisfiable by at least one cluster. For algorithms that require a status, we consider a state of having 95\% or greater utilization \emph{BUSY}. The status will change as work is assigned to a cluster; idle nodes will decrease, and queue depth will increase. For cost algorithms, we use per node costs from real data \cite{spheron_gpu_2026,umich_hpc_2026}. To mirror actual practices, we treat node allocations as exclusive.  For each simulation, we will randomly shuffle jobs and all satisfiable proposals, and run selection. A status of \emph{SELECTED} indicates a job  assigned to a cluster. In the case of not being satisfiable across contenders, it is considered \emph{REJECTED}. Any agent assessment with an \gls{api} error will be recorded as \emph{UNKNOWN}. The agent will always be provided with resource categories and providers discovered. After selection, the chosen computational resource would be dispatched.

\section{Results}
\label{sec:results}

We ran simulations to test our server components and agents against 200 simulated workers (75, 85, 40 for \emph{hpc}, \emph{cloud}, and \emph{standalone} archetypes, respectively) and 200 generated prompts for a total of 19,973 successful negotiation tests and 6,952  selections. Complete software, datasets, and analysis scripts are available \cite{vanessa_sochat_2026_19476827, mcpserver,resource-secretary, vanessa_sochat_2026_19476302, vanessa_sochat_2026_19476501}.

\begin{figure}[b!]
  \centering
  \includegraphics[width=\columnwidth]{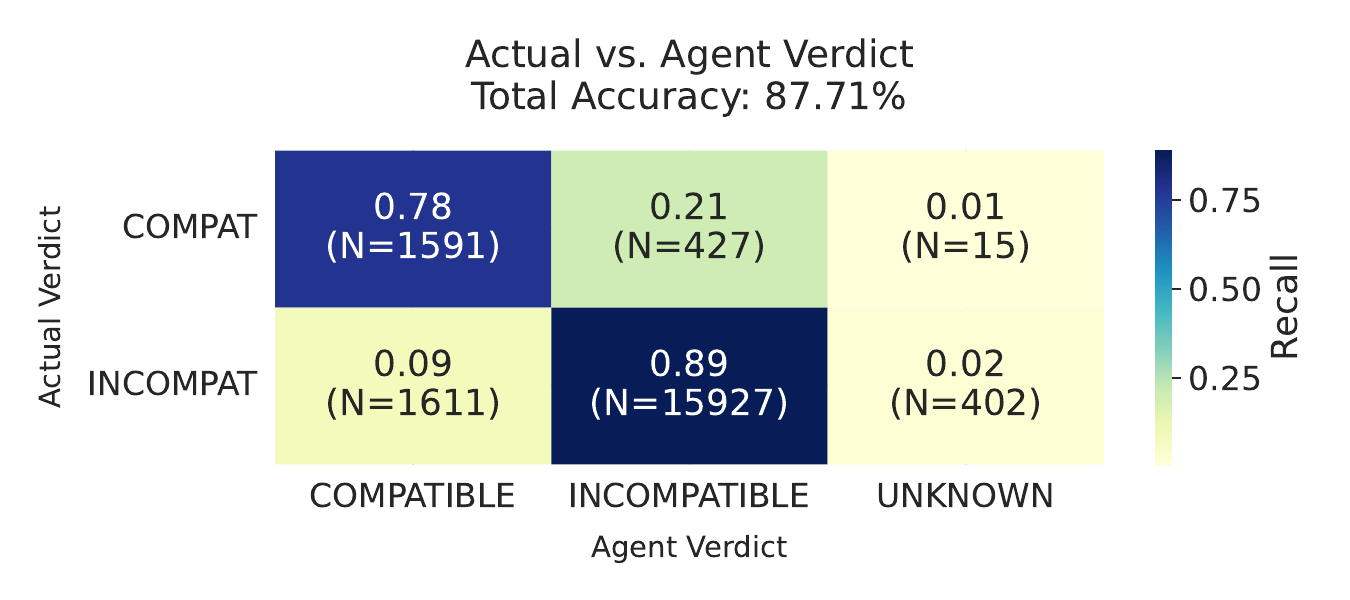}
  \vspace{-20pt}  
  \caption{\textbf{Worker truth versus agent verdict confusion matrix}. Normalization is done by row to account for class imbalances (N=19,973). Agents are skilled at detecting compatible and incompatible jobs. Accuracy is $(correct/total)$ requests}
  \label{fig:confusion-matrix}
\end{figure}

\smallskip
\noindent{\bf Reliability of Secretary Agent} 
\label{section:reliability-secretary-agent-results}
The degree to which an agent is able to correctly report the ability of a computational resource to satisfy a work request is shown in Figure \ref{fig:confusion-matrix}. The majority of assessments were correct (87.71\% overall accuracy), with the largest source of error resulting from an agent declaring an incompatible cluster compatible (N=1611).  Of these misses, we find that the agent missed the version requirement for software (N=1089), or a network (N=156), storage (N=180), or compute (N=186) requirement. There were 417 cases of \emph{UNKNOWN}, which resulted from the agent exceeding the maximum of 10 exploration loops allowed.

\begin{figure}[hb]
  \centering
  \includegraphics[width=\columnwidth]{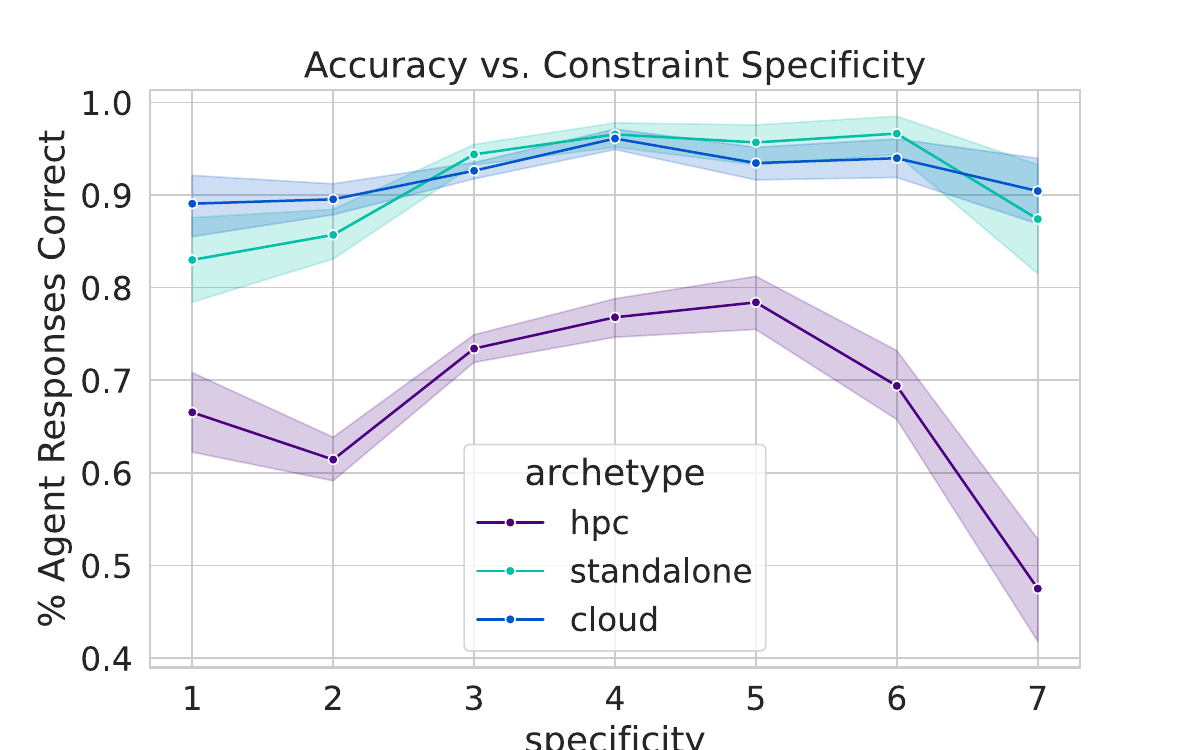}
  \caption{\textbf{Accuracy vs. Specificity}. Higher specificity indices correlate with an increased number of user requirements. Agents struggled with not enough information (1) or looking for a container technology (7).}
  \label{fig:accuracy}
\end{figure}

Figure \ref{fig:accuracy} shows the majority of error can be attributed to the \gls{hpc} archetype and specificity of 7, which explicitly modifies the request to add a container technology. From observation, we noticed that agents often strictly looked for a specific container technology, and could not deduce that different container technologies could be substituted. We also observed agents using the wrong tools to query for the installed container (e.g., calling a software tool instead of the container-specific tools) and determining the container was not found.

\begin{figure}[ht]
  \centering
  \includegraphics[width=\columnwidth]{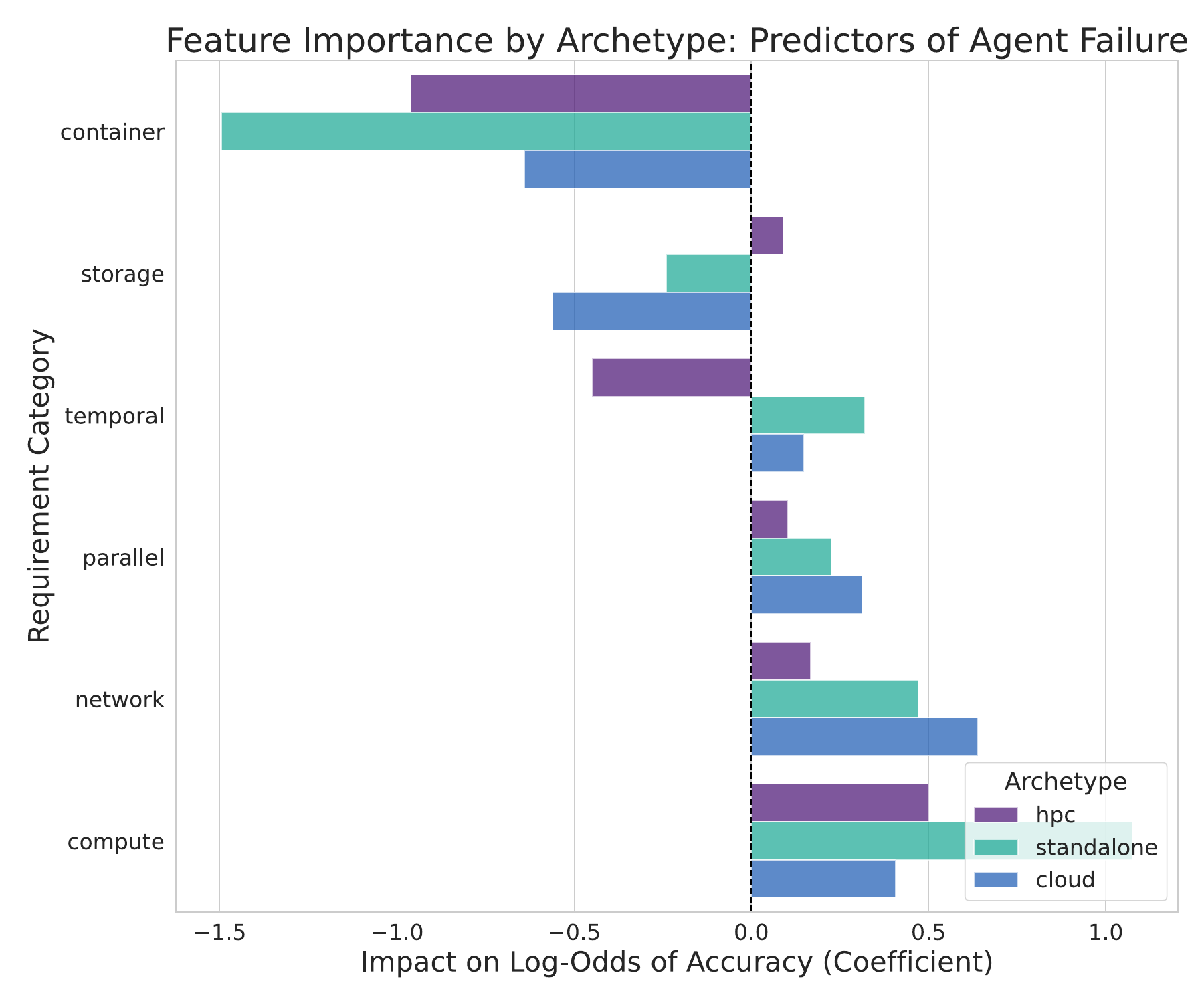}
  \vspace{-20pt}  
  \caption{\textbf{Beta values of resource categories for agent reasoning performance across archetypes}. Negatives values indicate features contributed to agent incorrect responses.}
  \label{fig:feature-influence}
\end{figure}

To determine categories that are associated with accuracy, we performed logistic regression from the Python \emph{statsmodel} package to calculate the log-odds impact, where a negative coefficient means that the category makes the task harder for the agent. Our logistic regression (Figure \ref{fig:feature-influence}) identifies several highly significant predictors of agent accuracy. For \emph{\gls{hpc}} workers, compute requirements significantly increase the log-odds of success ($\beta = 0.5$), while container requirements exert a strong negative impact ($\beta = -0.96$). For the \emph{standalone} archetype, compute is a positive driver ($\beta = 1.08$) and containers ($\beta = -0.96$) features negative. Finally, for the \emph{cloud} archetype compute ($\beta = 0.4$), network ($\beta = 0.63$), and parallel ($\beta = 0.31$) increase the log-odds of a correct verdict.

\begin{figure}[b!]
  \centering
  \includegraphics[width=\columnwidth]{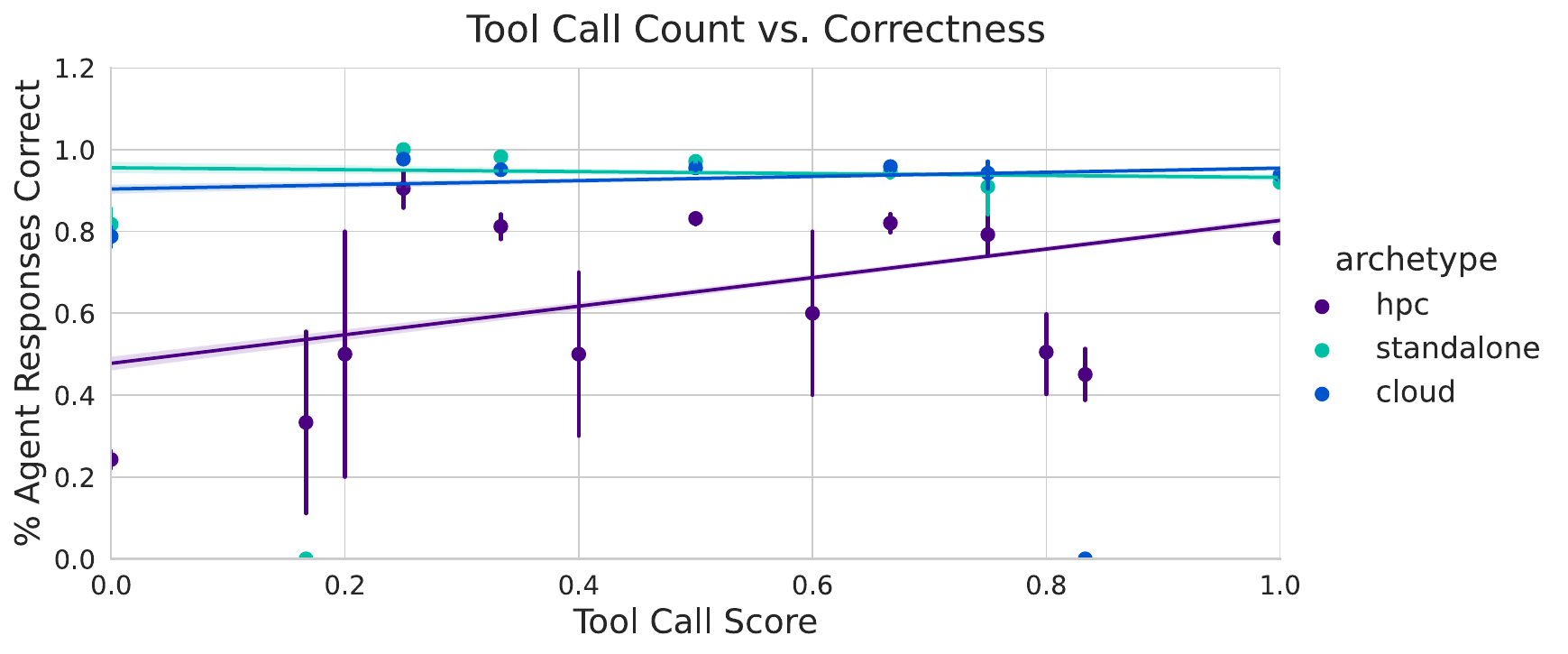}
  \vspace{-20pt}  
  \caption{\textbf{Accuracy of secretary agent responses versus tool call score}. The \gls{hpc} archetype had overall lower accuracy. More calls resulting in lower accuracy likely reflects trouble finding a needed requirement.}
  \label{fig:tool-calls}
\end{figure}

\smallskip
\noindent{\bf Tool Calls} 
\label{section:tool-calls}
We found no significant differences between an agent's ability to correctly predict or a machine's ability to satisfy a request. Figure \ref{fig:tool-calls} shows the percentage of agent responses correct versus the tool call score. A lower score (or score of 0) indicates that the agent failed to call tools that would be required to understand the environment. A tool call score of 0 does not indicate that the agent failed to call tools, but rather it failed to call any correct tools. Accuracy values above 80\% are notably present after approximately 30\% exploration of the space for all archetypes. With the exception of \gls{hpc}, more exploration does not necessarily lead to a better outcome. Interestingly, the \gls{hpc} archetype has a drop in overall accuracy for subsets of higher exploration.

\begin{figure}[t!]
  \centering
  \includegraphics[width=\columnwidth]{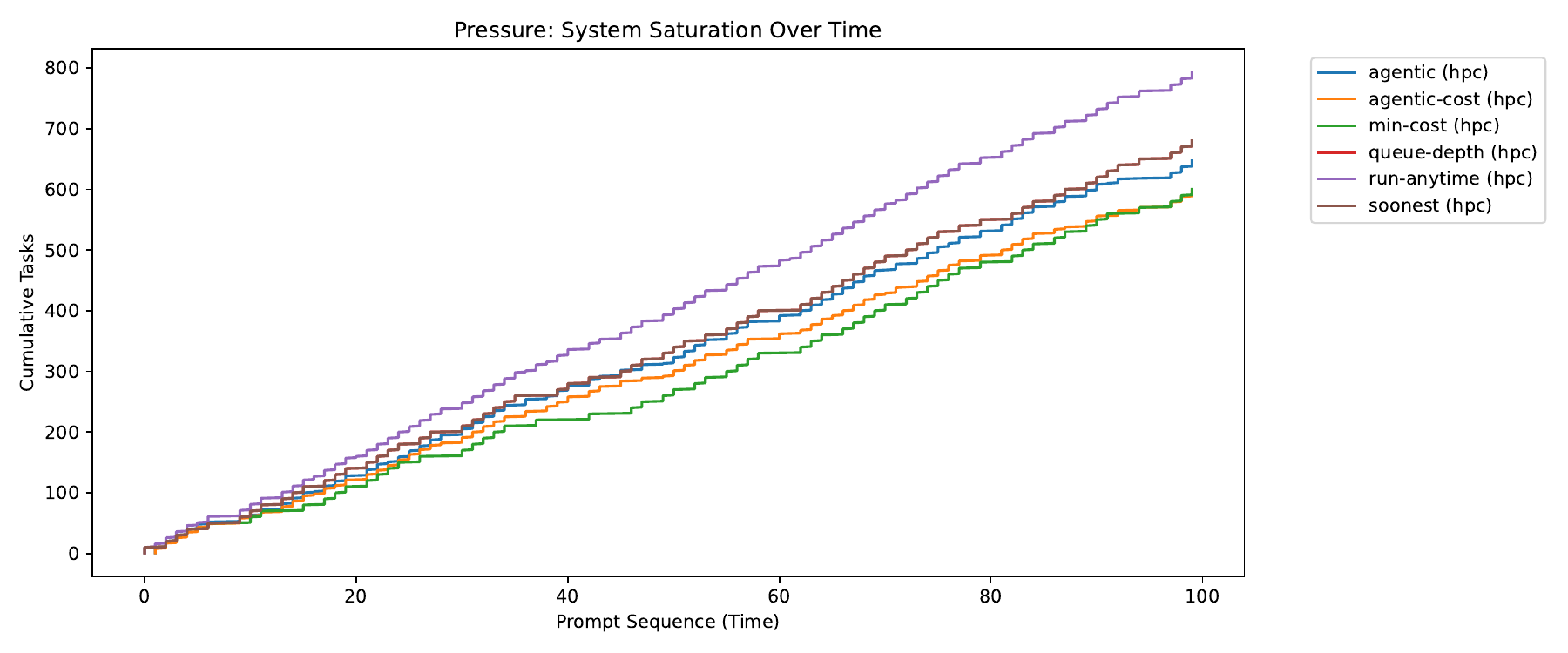}
  \vspace{-20pt}  
  \caption{\textbf{Selected worker pressure} as a sanity check for selection experiments. A single archetype \gls{hpc} with comparable queue sizes is shown.}
  \label{fig:pressure}
\end{figure}

\smallskip
\noindent{\bf Selection Simulations} 
\label{section:selection-simulations}
We evaluated the performance of agentic resource selection against traditional scheduling heuristics in a simulated environment consisting of a heterogeneous worker fleet (\emph{hpc}, \emph{cloud}, and \emph{standalone}). Traditional heuristics, including \emph{queue-depth}, \emph{min-cost}, and \emph{soonest} were compared to using a selection agent under two conditions -- with and without state. We found no differences between responses for with and without state, and report on with state (updating the queue) as a better representation of a true fleet. We ran 10 iterations of each of 8 strategies across 100 prompts for a total of 8000 potential results. Of that set, 6,952 simulated jobs were candidates for selection due to at least one compatible simulated worker, and 4,761 were successfully scheduled by algorithms given changing cluster state. We are primarily interested in the functionality of the simulated selection and software framework under realistic testing conditions. Toward this aim, we first did a sanity check to observe expected queue growth over time (Figure \ref{fig:pressure}) given queue depths reset to zero. These results were not used for subsequent analyses.

Agentic latencies for our main selection experiments were slower than more traditional heuristic strategies, an expected result since an agent needs to call a \gls{llm} \gls{api}. 
An agentic selection can take on the order of seconds, which inhibits high throughput. We find selection agreement between traditional approaches across 10 iterations (0.76), a likely reflection of the more deterministic nature of the algorithms. 

\begin{figure}[b!]
  \centering
  \includegraphics[width=\columnwidth]{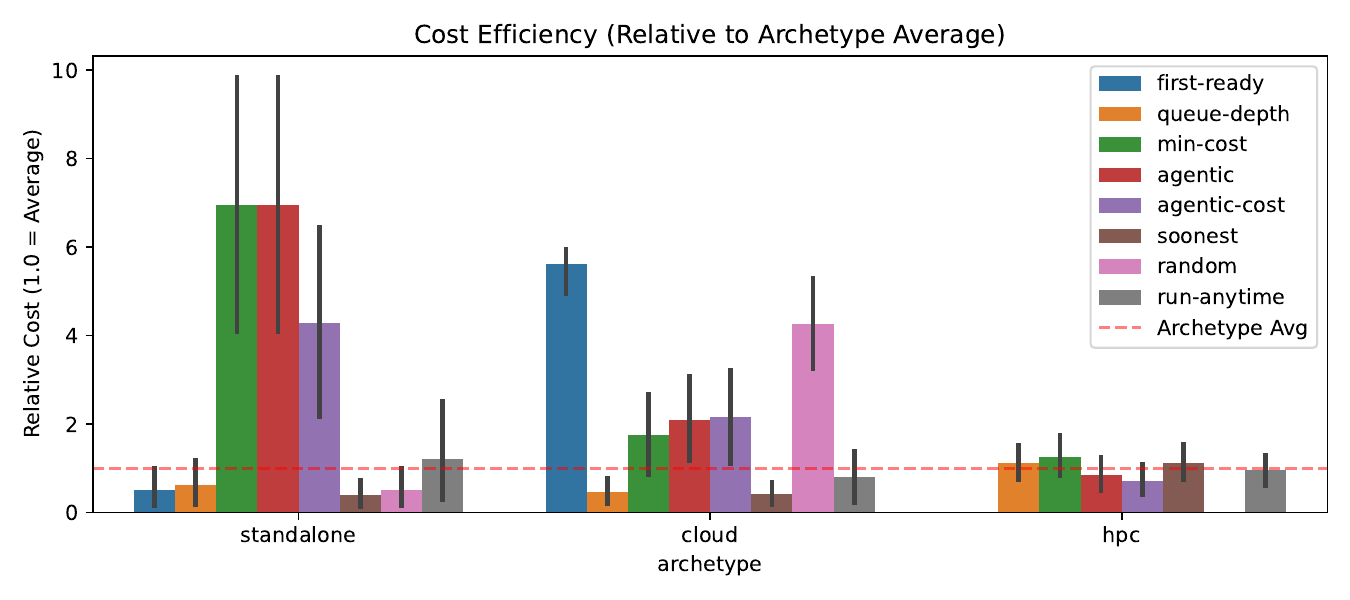}
  \vspace{-20pt}
  \caption{\textbf{Normalized cost efficiency between strategies}. The \emph{hpc} archetype did not get assignments for two strategies. Standalone costs were higher for traditional vs. agentic.}
  \label{fig:archetype-costs}
\end{figure}

We calculated a cost efficiency index (Figure \ref{fig:archetype-costs}) (CEI) to show cost normalized by archetype, as archetypes are designed to vary in scale and cost comparisons can be skewed. For the \gls{hpc} archetype, there are no significant differences in means. The \emph{first-ready} and \emph{random} strategies always selected a \emph{standalone} or \emph{cloud} instance, reflecting a likely greedy bias. As an example, for one simulation we see there are a total of 97 job prompts assessed. Of the 97, 18 are selected. Of the 18 selected, all are assigned to \emph{cloud} and \emph{standalone}. The \emph{agentic-cost} strategy outperformed its analogous manual component for the \emph{standalone} environment. Observation of the agent running revealed a tiered strategy of considering other resources provisioned by the node. Interestingly, we noticed the agent considering alternative resources in the case of an \emph{OR} gate for GPU or CPU. The agent often would select an allocation of a single node with the logic that the GPUs were more expensive, but the job would run faster. For our simulation we did not model application running time. More work is needed to add this to our model. The increased cost of a GPU node over a CPU node does not necessarily translate to improved application performance.  The \emph{cloud} archetype demonstrates the lack of informed choice for the \emph{first-ready} and \emph{random} strategies.

\section{Discussion}
\label{sec:discussion}

\smallskip
\noindent{\bf Reliability of Secretary Agent} 
\label{section:reliability-secretary-agent-discussion}
In our work we observe that the secretary agent is skilled at determining a computational environment is not compatible for a workload. We believe this results from the task itself -- it is easier to rule out an environment due to a missing dependency than verify that every detail is correct. The highest risk case for job negotiation occurs when an agent determines that a computational resource is compatible for an environment when it is not, representing 1,611 of our test cases. Closer inspection told us that of the large portion that incorrectly identified matching software (N=1089) the agent would make a call to a manager that returned dependencies (e.g., \emph{spack find}) and mix up the version of the required software with a dependency. We hypothesize this is fixable by making tool call metadata more specific and decisive. Unlike human output of the command that is verbose with dependency versions, an agent should only be provided with the exact package metadata it is looking for, and a separate tool call might be exposed to relay dependency information. 
This finding suggests that tool call interfaces are important, and cannot always exactly mirror the human equivalent. 

We found that the majority (62.5\%) of cases that were false negatives (stated not compatible when they were) were of specificity levels 1 and 7, where the user requested an application name only. For a level of 1 that only provides the application without version to run, secretary agents did not have enough information. For cases of level 7, the agents had trouble finding the container software. A second error that the agent made was to try to identify the network type by inspecting package managers. 


\noindent{\bf Limitations} 
\label{section:limitations}
Our simulated archetypes are a best effort to generate a cohesive set of reasonable environments. They do not (and cannot) reflect every real-world environment. We want to improve upon the design of an archetype to better capture more environment dimensions, such as including multi-tenancy, a count of users, and security or workload manager policies. When we add a scheduling component, we plan to simulate applications and more fine-grained temporal requirements. A more realistic query might require a workload to be done within a specific number of hours, or linked in time with job dependents.

Probing is limited to where the worker is running. In practice, it would run on a login node or exposed worker node of a job. In the case of a launcher node, it needs to be the case that the launcher node uses tools that give access to full cluster resources. Generally speaking, it is up to the cluster or single server resource to expose the minimum set of providers needed. The agent also needs credentials for a \gls{llm}, and there is a tradeoff between cost paired with performance, and \gls{llm} time to respond that needs to be better studied. Results would vary based on the model used. 

Our selection experiments were primarily done to test the mcp-server's ability to orchestrate them, and more sophisticated algorithms and experiments are warranted. 
The benefit of using a selection agent over a hard-coded algorithm is that it is flexible to uncertainty. It is flexible to receiving different preferences and policies on the fly, or even an auction-based approach for agents to bid for work. We believe that traditional algorithms can be combined with agents to enable better decision making. Agents can be given algorithms as tools to call to make a choice. Our selection simulation does not account for job shape (request topology) or the idea that the same work can be satisfied by different shapes. Shape has implications for scheduling and performance and work we are actively doing integrated with the Flux Framework workload manager. This is tied to the idea that a secretary agent could return a modified suggestion for another shape of resources. We are actively working on several of these ideas.

Our work is also limited in that we have not addressed authentication. Our experiments currently assume that a deployment is owned by a single user or group. We expect that some of the limitations and errors that the agents made are due to the older version of the models we used (Gemini 2.5). There is a tradeoff between performance with cost, and time. Better performing agents often take longer to run and are more expensive. Agents are evolving so rapidly that it is challenging to perform experiments and publish work that represents a latest model. Many of these limitations may be mitigated by using the latest models. 



\smallskip
\noindent{\bf Future Work} 
\label{section:future-work}
We plan to continue work testing agentic approaches for server hierarchies and component designs. 
For example, a middle level mcp-server might serve as a hub and a worker, or a cluster might provision multiple hubs across the top level using the RAFT consensus protocol \cite{dautov2025controlled}. An mcp-server also requires different policies for when to automatically de-register or not trust a worker. 

While we did not take advantage of events in our experiments, delivering events back to a hub and calling client to update dispatch state is future work. Updating state would be important if the hub serves as a central job manager. 
While the mcp-server receiving function \emph{ask\_secretary} is currently provisioning agents that use discovered resource providers, this is not a hard requirement for the future of the resource secretary project. An equivalent, more deterministic algorithm can be substituted, and even one that uses the same provider interfaces to get information about the cluster. 

Notably, we did not include dispatch simulations and results in this work, and we have completed work for this case. 

\section{Conclusion}
Agentic frameworks are a central component of the future for not just the \gls{hpc} community, but for software in general. Autonomous, converged \gls{hpc} infrastructure that can intelligently negotiate job workloads is becoming feasible.  From autonomous task execution to failure recovery and AI-supported scheduling techniques, these powerful transitions have and will continue to change the day to day work of computational scientists and engineers. We anticipate a new era of faster, more accurate, and more accessible computing for the next generation of \gls{hpc}.

 \section*{Acknowledgment}
 Thank you to olive oil, quiet sunsets, and feel of the crisp air on the first
 steps out into the cold. Thank you to our cloud collaborators for continued
 interesting discussion in this area. Thank you to Livermore Computing for
 supporting me in all respects (VS).

 This work was performed under the auspices of the U.S. Department of Energy by
 Lawrence Livermore National Laboratory under Contract DE-AC52-07NA27344 and
 was supported by the LLNL-LDRD Program under Projects No. 24-SI-005
 (LLNL-CONF-2018020).

\bibliographystyle{IEEEtran}
\bibliography{references}

\end{document}